# INTEGRATING UNIVERSAL GENERATIVE AI PLATFORMS IN EDUCATIONAL LABS TO FOSTER CRITICAL THINKING AND DIGITAL LITERACY

Vasiliy Znamenskiy, Rafael Niyazov, and Joel Hernandez

Science Department, Borough of Manhattan Community College, The City University of New York, New York, USA

## *ABSTRACT*

*This study investigates the educational potential of generative artificial intelligence (GenAI) platforms based on large language models (LLMs), such as ChatGPT, Claude, and Gemini, as tools for student-centred learning. Recognizing the current limitations of GenAI, particularly its propensity for generating inaccurate or misleading information—the paper proposes a novel instructional strategy: an interdisciplinary laboratory designed to foster critical evaluation of GenAI-generated outputs.*

*In this pedagogical model, students engage with GenAI systems by posing questions or solving problems drawn from topics they have already studied and understand. Equipped with correct answers, students are positioned to assess the accuracy and relevance of AI-generated responses across multiple modalities, including text, images, and video.*

*Students design challenging prompts and tasks that help compare the intellectual performance of various GenAI. This approach was implemented for a specially designed lab session within a general astronomy course for non-science majors. Multiple student groups completed the lab, demonstrating high levels of engagement, initiative, and critical thinking. Findings suggest that such activities not only deepen students' comprehension of the scientific content of learning courses but also cultivate essential skills in digital literacy and critical interaction with AI technologies.*

## *KEYWORDS*

*Generative Artificial Intelligence, GenAI, Large Language Models, LLMs, Critical Thinking, Digital Literacy, AI in Education, ChatGPT, Claude, Gemini, AI-augmented Learning, Interdisciplinary Education*

## 1. INTRODUCTION

The current phase in the evolution of artificial intelligence is marked by the rapid advancement and widespread adoption of large language models (LLMs), such as ChatGPT (OpenAI), Gemini (Google), and Claude (Anthropic) [1]. While earlier dialogue systems like ELIZA and ALICE laid the foundational groundwork, recent progress in transformer architectures, deep learning, and cloud-based computing [2] has resulted in a significant leap in the performance and generative capabilities of modern systems. These advances have expanded the functionality of LLMs from basic language comprehension to the generation of coherent text, executable code, images, and video content.

Contemporary generative AI (GenAI) dialogue systems exhibit notable proficiency in processing open-ended user inputs and generating diverse outputs, including written texts, multimedia materials, and programming scripts. Their rapid integration into everyday life has sparked widespread public debate, bringing AI-related discussions beyond technical and scientific contexts [3] into educational, ethical, and sociopolitical domains.

In education, the rise of GenAI technologies has elicited mixed responses. On the one hand, these tools promise to support personalized learning, enhance student motivation, and facilitate more effective pedagogical interactions [4][5] through intelligent tutoring systems and tailored educational chatbots. On the other hand, concerns are growing about the erosion of academic integrity [5][6], as students increasingly rely on GenAI tools for completing assessments and assignments without genuine engagement, potentially leading to superficial learning and diminished cognitive development.

Developing competencies for the responsible and informed use of GenAI [2][7] is becoming essential for 21st-century graduates. Within the broader agenda of digital transformation in higher education [3], institutions bear a responsibility to promote the ethical deployment of AI for societal benefit. This includes embedding critical digital literacy into curricula [3][6] and offering systematic training for both students and faculty on the legal, ethical, and data security dimensions of AI use. Such efforts support not only individual academic and professional growth but also the formation of a more inclusive and equitable digital society.

Despite the potential of GenAI, intelligent educational assistants and domain-specific chatbots [8][9] are not yet widely used in postsecondary teaching practices. However, general-purpose LLMs are increasingly being adopted by students, often as tools for generating assignment responses with minimal engagement in the learning process. This trend necessitates a critical reassessment by educators of their pedagogical stance toward AI usage in academic work.

One response has been to limit digital access in classrooms through bans on smartphones or restricted internet connectivity. An alternative and more constructive strategy is to incorporate chatbot interactions into formal instruction [4]. In this model, educators serve as guides and facilitators, helping students learn to engage with GenAI systems critically and effectively. Given that LLMs, while powerful, are prone to producing plausible but erroneous or misleading outputs [6][9] —sometimes in unexpectedly straightforward contexts—students must be equipped not only with the ability to formulate effective prompts but also to critically evaluate AI-generated information.

Consequently, there is an emerging need to design interdisciplinary educational tasks [7][9] that simultaneously support subject-matter learning, cultivate critical thinking, and foster responsible AI engagement skills.

While banning or limiting access to AI may seem like a straightforward solution, it risks missing a deeper opportunity: to treat AI not merely as a tool to regulate but as a subject of critical learning.

Generative artificial intelligence (AI) tools, such as ChatGPT, Claude, and Google's Gemini, have rapidly become commonplace in daily life and are now permeating educational settings. Many students already incorporate these systems into their study routines. However, they often use them merely to expedite their assignments, accepting AI-generated answers at face value without critical examination. This uncritical reliance on AI can lead to superficial learning outcomes and a lack of deep understanding of the subject matter.

As a result, educators face a critical challenge: how to foster thoughtful and meaningful student engagement with these tools. Rather than banning these systems or ignoring their influence, a more constructive strategy is to integrate AI into the learning process as an object of critical inquiry. Students should be encouraged to actively test and evaluate AI-generated responses. For instance, they might cross-check the AI's outputs against reliable sources or their own knowledge and reflect on the AI's performance (considering its accuracy, biases, and limitations). Through this process, generative AI tools can be transformed from mere conveniences into catalysts for deeper learning. Such an approach has the potential to sharpen students' critical thinking skills, deepen their comprehension of academic content, and cultivate a greater understanding of the technologies that are increasingly shaping their education.

## 2. STUDY AIM AND SCOPE

This study investigates the educational potential of contemporary chatbots, with a focus on their application in both direct student engagement and as support tools for instructors, particularly in tasks such as generating assignments, designing assessments, and evaluating student submissions. Given the rapid advancement in language models and digital technologies, it is plausible to hypothesize that future developments may enable general-purpose chatbots to assume responsibilities across the entire educational cycle. Such a vision envisions an advanced pedagogical AI system capable of:

- Designing curriculum frameworks and course structures.
- Delivering instructional content interactively.
- Facilitating the development of problem-solving abilities and addressing conceptual queries.
- Monitoring and assessing student progress through formative and summative evaluations.
- Generating final grades by integration with centralized academic records systems.

This leads to the conceptualization of a theoretical "ideal chatbot"—a fully autonomous digital educator capable of replicating the core functions of human teaching professionals across all stages of instruction.

However, despite significant progress in generative AI, particularly in natural language understanding and the creation of textual, visual, and interactive educational resources, current chatbot technologies fall considerably short of this ideal. Notable limitations include:

- Inability to perform long-term planning aligned with academic schedules.
- Absence of multi-user or group-based communication frameworks within the chatbot interface.
- Limited capacity for contextual pedagogical adaptation, including real-time responsiveness to individual learner needs.
- Weaknesses in interpreting ambiguous tasks and a lack of mechanisms for accountable error recognition.
- Deficiencies in ethical and interpersonal decision-making—key dimensions of effective teaching practice.

This study aims to explore the current capabilities and limitations of general-purpose chatbots through practical experimentation, providing a systematic evaluation of their suitability for educational use. The analysis identifies both the technological constraints and the pedagogical risks associated with their deployment and proposes recommendations for their ethical and meaningful integration into contemporary learning environments.

Preliminary findings suggest that, at present, it is premature to recommend chatbots for autonomous use by students, even as personal learning assistants. Despite their sophisticated linguistic capabilities, these systems remain prone to inaccuracies, including the phenomenon of "AI hallucinations," whereby outputs may appear plausible yet are factually incorrect, often in critical or high-stakes contexts.

The study also briefly examined the potential role of chatbots in automating instructional design tasks, such as generating quizzes, learning materials, and examination content. These outputs could, in principle, be exported into formats compatible with widely used learning management systems such as Blackboard or Brightspace [10]. Nevertheless, this paper primarily concentrates on the implications of student–chatbot interaction, rather than instructor-focused applications.

## 3. TRANSFORMING LIMITATIONS INTO LEARNING: CRITICAL EVALUATION OF GENERATIVE AI OUTPUTS IN INTERDISCIPLINARY LAB WORK

Considering the persistent error rates exhibited by large language models—rates that remain considerable for reliable use in formal educational settings—this study proposes reframing such limitations as pedagogical opportunities. Specifically, the development of laboratory-based activities is suggested, wherein students critically examine chatbot-generated outputs, including written responses, explanations, and problem solutions. By identifying inaccuracies and assigning evaluative scores to these outputs, students not only refine their critical thinking skills but also achieve a more nuanced and thorough grasp of the subject content.

Expanding on this approach, the paper introduces the concept of a novel interdisciplinary instructional format: the *Laboratory on Critical Evaluation of GenAI Outputs*. This model shifts the emphasis from the use of AI as a passive instructional tool to its role in cultivating essential academic competencies. These include scientific rigor, analytical reasoning, and the capacity to navigate and assess potentially unreliable or misleading information—skills increasingly vital in today's AI-augmented learning environments.

## 4. PEDAGOGICAL DESIGN OF A LABORATORY ON EVALUATING GENERATIVE AI IN SUBJECT-SPECIFIC CONTEXTS

A defining characteristic of the proposed laboratory exercise is its implementation at the culmination of a course, when students have already acquired foundational knowledge and core competencies in the subject area. The primary aim is to assess the performance of various generative AI chatbots in relation to the specific content of the course, encouraging students to adopt a critical stance toward the quality and reliability of AI-generated information.

This laboratory task centers on active student engagement in the construction and analysis of prompts directed at general-purpose language models. Students are required to independently formulate a set of questions that span essential aspects of the course content and submit these to multiple chatbot systems. The dual purpose of this exercise is to assess students' mastery of the material and to cultivate their ability to evaluate the interpretative capabilities and limitations of GenAI tools within a disciplinary framework.

Crucially, students are guided to develop questions that differentiate chatbot responses by degrees of correctness, ranging from wholly accurate to partially accurate to clearly erroneous. To achieve this, students must:

- Review and internalize previously covered course material.
- Design questions of varying complexity and specificity.
- Develop evaluative criteria to assess the quality of chatbot-generated responses.

The task must be carefully calibrated: questions should avoid being overly simplistic (which may result in uniformly correct responses from all systems) while also steering clear of highly ambiguous or excessively specialized formulations that might falsely suggest chatbot inadequacy. Upon collecting responses, students analyze the output for accuracy, completeness, and conceptual depth. Responses are categorized into three types:

1. **Fully correct** – responses that are factually accurate, logically coherent, and well-articulated.
2. **Partially correct** – responses that exhibit minor omissions, reasoning gaps, or stylistic flaws;
3. **Incorrect** – responses containing factual inaccuracies, contradictions, or misleading interpretations.

To support a structured evaluation, students apply a multi-dimensional rubric, which may include criteria such as:

- Terminological precision.
- Contextual appropriateness.
- Depth of explanation.
- Logical coherence.

Through this analytical process, students develop critical thinking skills, reinforce their understanding of course content, and gain practical experience interacting with GenAI tools. Moreover, the activity fosters an interdisciplinary perspective on the role of AI in education by highlighting both the capabilities and limitations of language models in content-specific applications.

The exercise concludes with a comparative ranking of chatbot performance based on the aggregated scores assigned by students, offering an evidence-based assessment of their relative competence within the course domain.

## 5. Expanding Interdisciplinary GenAI Laboratory Work: A Pilot Implementation in Astronomy Education

We propose the development of interdisciplinary laboratory activities that leverage generative artificial intelligence (GenAI) across a range of academic disciplines. As an initial implementation, this pedagogical model has been introduced within a general astronomy course designed for students from non-STEM backgrounds.

The structure of the laboratory work may take the form of a single, integrated assignment with three interconnected components—text, image, and video—or as three discrete tasks, each dedicated to one modality. This tripartite framework is intended to acquaint students with a variety of generative tools, enabling them to interact not only with language-based models but also with systems capable of producing visual and multimedia content.

Astronomy was selected for this pilot due to its inherently imaginative and open-ended nature, which makes it particularly conducive to creative engagement with generative technologies. Unlike more rigidly structured scientific disciplines, astronomy invites speculative exploration. GenAI tools—ranging from textual to visual chatbots—allow students to construct original space-related narratives, conceptualize hypothetical celestial phenomena, and generate fictional cosmological scenarios without requiring strict adherence to current scientific paradigms. This flexibility lowers the initial cognitive barriers associated with using GenAI, fostering greater student enthusiasm and creative experimentation.

The approach supports multiple pedagogical objectives simultaneously: enhancing students' digital and GenAI literacies, promoting interdisciplinary thinking, and cultivating competencies in visual communication and critical information analysis. Furthermore, it helps students reconceptualize AI not merely as an automation tool but as an active partner in both cognitive and creative processes.

Looking ahead, this methodology will be extended to other disciplines, including physics, chemistry, and biology, with appropriate modifications to reflect the epistemological and instructional characteristics of each field. This expansion aligns with a broader vision of transdisciplinary digital transformation in higher education, in which GenAI tools are integrated thoughtfully into domain-specific learning processes.

## 6. IMPLEMENTATION OUTCOMES AND PEDAGOGICAL IMPACT OF AN AI-INTEGRATED ASTRONOMY LABORATORY FOR NON-SCIENCE MAJORS

As part of an experimental educational initiative, a laboratory activity was designed and piloted within the *General Astronomy* course for non-science students. The instructional materials included a comprehensive methodological guide, detailed instructions for conducting the lab, and integrated links to digital forms for documenting student results.

The laboratory activity was implemented across several student cohorts over multiple academic semesters. Each session was allotted two academic hours. While this duration proved insufficient for all students to complete all three components of the lab (text, image, and video tasks), a notable number of participants demonstrated rapid comprehension and completed the full scope of the assignment within the allocated time.

Particularly noteworthy were the outcomes from the second and third phases of the lab, which involved generating visual representations of imagined astronomical objects and phenomena using GenAI tools. These images and videos—interpreted as speculative extraterrestrial scenarios—were showcased during class sessions and stimulated vibrant peer discussion. This not only fostered heightened cognitive engagement but also promoted collaborative learning and creativity.

Students who were unable to complete the full task during class were allowed to finalize their projects independently. Final submissions, particularly images and videos, were voluntarily uploaded to a dedicated Facebook group for the course (https://www.facebook.com/groups/GeneralAstronomy). This online platform functioned as a supplementary engagement space, enhancing student participation and extending the learning community beyond the classroom. Many students exceeded the minimum requirements of the assignment, displaying substantial initiative and academic curiosity.

Moreover, several individual projects produced through this laboratory activity were subsequently adapted into poster presentations for the annual BMCC student research symposium (BARS). Student participation in the symposium was marked by a high degree of motivation and received positive feedback from both peers and faculty, further underscoring the educational value of integrating generative AI into the learning experience.

# 7. AI-Enhanced Astronomy Lab Manual

## Introduction: Artificial Intelligence in Astronomy

Artificial intelligence (AI) is playing an increasingly transformative role in the field of astronomy and space exploration. Key applications include:

- Data Processing and Image Enhancement: Machine learning techniques are employed to refine astronomical images from space telescopes (e.g., Hubble, James Webb), eliminating noise and increasing clarity. Notably, AI was instrumental in enhancing the historic image of the supermassive black hole in galaxy M87.
- Pattern Recognition and Classification: AI systems assist in categorizing galaxies, stars, and exoplanets, improving both the efficiency and precision of classification tasks. These systems can also detect patterns across vast astronomical datasets, enabling the discovery of previously unknown phenomena.
- Spacecraft Autonomy and Hazard Prediction: AI contributes to spacecraft autonomy by processing environmental data and predicting threats such as solar flares and debris fields. These capabilities optimize mission planning and reduce dependence on ground control.
- Forecasting Cosmological Events: AI models trained on historical data can anticipate rare events, such as gravitational wave emissions and supernovae, thereby guiding observational efforts more effectively.
- Scientific Efficiency and Collaboration: By automating repetitive analytical processes, AI enables scientists to focus on higher-level reasoning and hypothesis testing, accelerating research and reducing operational costs.

As AI technologies continue to advance, their integration into astronomy is expected to drive further scientific breakthroughs. In this laboratory module, students will explore how generative AI (GenAI) can enhance understanding of key concepts in astronomy, cosmology, astrophysics, astrobiology, and space exploration.

## Defining Generative AI in an Educational Context

Generative AI refers to systems that autonomously produce original content—text, images, video, audio, or code—based on user-defined prompts. These systems are powered by advanced machine learning architectures such as large language models (e.g., GPT), generative adversarial networks (GANs), diffusion models, and transformers. The use of GenAI tools in education presents novel opportunities to simulate phenomena, process scientific data, and encourage creative inquiry.

## Lab Objectives and Educational Aims

The lab is designed to achieve the following:

- Understanding GenAI Capabilities: Enable students to experiment with and evaluate various GenAI platforms in the context of astronomy.
- Developing Critical Analysis Skills: Encourage students to assess the reliability, consistency, and limitations of AI-generated outputs across different systems.
- Enhancing Engagement: Foster active student participation through practical and creative interaction with digital tools.
- Building Technological Proficiency: Cultivate essential skills in navigating AI tools increasingly relevant to modern scientific research and academic workflows.

**Expectations and Learning Approach**

- **Active Engagement**: Students are expected to interact thoughtfully with GenAI systems, completing each task attentively and reflectively.
- **Critical Inquiry**: Students should approach each activity analytically—interrogating AI responses, identifying errors, and drawing reasoned conclusions.
- **Collaborative Learning**: Peer discussion is encouraged to build a shared understanding of GenAI performance and facilitate collective problem-solving.
- **Innovative Thinking**: Students are invited to devise challenging, imaginative prompts that test the limits of GenAI reasoning and enhance conceptual understanding.

**Lab Components**

**1. Evaluating GenAI Text Platforms**

**Objective**: Students will design astronomy-related questions, input them into multiple GenAI chat platforms, and compare the accuracy and clarity of the responses.

**Expected Learning Outcomes**:

- Ability to craft precise, content-relevant questions.
- Comparative analysis of GenAI-generated answers across platforms.
- Increased familiarity with various AI tools for scientific inquiry.

**Required Materials**:

- Computer with Internet access
- Lab Report Template: Google Doc Link
- GenAI Platforms:
  - ChatGPT
  - Gemini
  - Claude
  - Microsoft Copilot (Bing)

**Procedure**:

- Devise ten astronomy-themed questions with known answers (e.g., from textbooks or course materials).
- Input questions into each GenAI platform and evaluate the responses.
- Attempt to identify questions where different platforms yield divergent or incorrect answers.
- Analyse performance by noting which platform produces the most accurate and helpful responses.

**Example Topics**: Planetary order, celestial mechanics, notable space missions, moon phases, stellar evolution, and observational astronomy.

**2. Creating AI-Generated Astronomical Imagery**

**Objective**: Students will explore GenAI image generators to visualize cosmic phenomena and assess the realism and creativity of the generated outputs.

**Expected Learning Outcomes**:

- Develop prompt engineering skills tailored to scientific visualization.
- Compare GenAI-generated images with reference astronomical imagery.
- Understand the creative liberties and scientific constraints of AI-based image generation.

**Recommended Platforms**:

- Night Cafe
- Bing Image Creator
- Google ImageFX
- Craiyon

**Procedure**:

- Create detailed prompts describing astronomical objects or events.
- Generate images using multiple platforms and compare outputs.
- Assess images based on scientific plausibility, visual fidelity, and imaginative appeal.
- Include images and descriptive prompts in the final lab report.

**3. Visualizing Dynamic Astronomical Phenomena via GenAI Video Tools**

**Objective**: Students will produce short educational videos illustrating dynamic space phenomena (e.g., black hole mergers, eclipses, planetary motion) using AI video platforms.

**Expected Learning Outcomes**:

- Ability to research and storyboard complex phenomena.
- Experience with GenAI video tools to communicate scientific content effectively.
- Improved science communication skills through multimedia storytelling.

**Suggested Tools**:

- InVideo
- Kapwing
- Synthesia

**Procedure**:

- Select a phenomenon and conduct brief background research.
- Write a concise script or storyboard outlining the visual sequence.
- Use GenAI video platforms to generate the visual content, adding narration or captions if desired.
- Submit completed videos via Blackboard and provide a brief reflection on the creative and technical process in the lab report.

**Conclusion**

This AI-enhanced astronomy lab empowers students to critically engage with generative technologies, deepen their understanding of scientific concepts, and develop skills relevant to the future of interdisciplinary education. By exploring AI's strengths and limitations across textual, visual, and audiovisual outputs, students are not only positioned as consumers of educational content but also as informed and reflective creators who can leverage GenAI tools in their academic and professional trajectories.

## 8. DISCUSSION

This project demonstrated that the integration of generative artificial intelligence (GenAI) tools into laboratory-based instruction can significantly enhance student engagement, particularly among learners from non-scientific academic backgrounds. The lab not only supported content acquisition in astronomy but also encouraged the development of critical inquiry, digital literacy, and confidence in interacting with emerging technologies.

### 8.1. Strengths and Pedagogical Benefits

Student engagement throughout the lab activities was notably high. Learners responded positively to the opportunity to interact with various GenAI platforms, using them to generate text, visualizations, and video content. These multimodal tasks fostered creativity and enabled students to explore complex scientific themes in a manner that was both accessible and personally meaningful.

The comparative analysis of outputs from different GenAI tools further deepened students' understanding of how these systems function, highlighting both their generative potential and their limitations. The voluntary continuation of projects beyond the classroom, including sharing outputs on digital platforms and presenting at student symposia, indicates a high level of intrinsic motivation and meaningful engagement. Such active learning experiences are widely recognized as more effective than passive instructional formats such as lectures or readings.

### 8.2. Challenges and Considerations

Despite these successes, the implementation of GenAI in education raises several challenges. One of the most pressing concerns is the potential for students to accept AI-generated responses uncritically. GenAI systems, such as ChatGPT, may produce responses that appear coherent and plausible but are factually inaccurate or misleading. Without sufficient guidance, students may internalize incorrect information, compromising their understanding of scientific concepts.

It is therefore essential that educators provide explicit instruction in critical evaluation skills and encourage students to verify AI-generated content against authoritative sources. Moreover, students must be taught to understand that GenAI systems do not possess knowledge in a human sense but operate probabilistically, generating responses based on statistical patterns in language data rather than verified facts.

### 8.3. Toward More Informed and Reflective Use of GenAI

The lab experience contributed to a pedagogical shift in students' approach to AI tools—from reliance on GenAI for quick answers to a more reflective and analytical engagement. Students learned to differentiate between high-quality, incomplete, and erroneous responses and developed skills in crafting effective prompts and evaluating the educational value of AI outputs.

Overall, the project underscores the potential of GenAI to serve as a constructive educational resource when integrated thoughtfully and supported by appropriate scaffolding. With educator guidance, GenAI can move beyond being a passive content generator to becoming a catalyst for deeper thinking, creativity, and digital competence in 21st-century learners.

### 8.4. Rethinking the Limitations of the Study Considering AI-Student Interaction

If we judge this study by the usual rules of academic research, it might seem weak. There was no control group, no standardized test to measure learning, and no way to repeat the same experiment. But these aren't just mistakes — they reflect a new and unusual way for students to interact with AI.

Generative AI tools like ChatGPT don't always give the same answer to the same question. They answer differently depending on how and when the question is asked. Because of this, using rigorous research methods doesn't always make sense. Trying to control everything would deprive students of the opportunity to explore and learn naturally.

Instead, this study allowed students to freely try things out and learn by doing. And it revealed some important things: students learned to ask better questions, check the AI's answers, and think more deeply when the AI made mistakes. They even continued to work with the AI after the session ended, which shows real interest and curiosity.

To better understand this type of learning in the future, we need new ways to study it. We need to look at how students use AI over time, what questions they ask, how they respond to incorrect answers, and how their thinking changes. Instead of focusing solely on test scores, we need to look at how students develop their ability to use and question AI effectively and intelligently.

In short, while this study may not follow all the usual rules of educational research, it opens the door to a better understanding of how students can learn and teachers can teach in today's world, using often-obscure AI as a tool and as a learning partner.

## 9. DIFFERENTIATING THE EDUCATIONAL UTILITY OF TEXT, IMAGE, AND VIDEO-BASED GENERATIVE AI TOOLS

Our evaluation of currently available generative artificial intelligence (GenAI) platforms indicates notable differences in their educational applicability, depending on the modality of content generation—text, image, or video.

**Text-based GenAI tools**, such as ChatGPT and Claude, have demonstrated considerable pedagogical value. These systems excel in delivering detailed explanations, structured reasoning, and context-rich narratives that support the comprehension of both foundational and advanced theoretical concepts. When tested against questions of typical undergraduate textbook difficulty, text-based GenAI systems produced accurate responses in over 90% of cases. However, occasional inaccuracies—often presented in a confident and linguistically plausible manner—highlight the need for critical oversight. The conversational interface of these tools fosters student engagement by enabling iterative questioning and immediate clarification, while the ability to compare outputs from different platforms supports the cultivation of critical evaluation skills.

**Visual GenAI tools**, including platforms such as NightCafe and Craiyon, offer the potential to enhance learning through visual representations, which are often more memorable and accessible than textual explanations. Nonetheless, at their current stage of development, these platforms exhibit significant limitations in scientific accuracy. Empirical testing indicates that graphical GenAI systems fail to produce reliable illustrations or diagrams in response to prompts involving college-level scientific or technical content, with error rates approaching 100% in such contexts.

**Video-based GenAI systems**, such as Synthesia and InVideo, are similarly promising for depicting dynamic processes and enhancing engagement through audiovisual storytelling. However, their outputs currently lack the scientific precision necessary for accurate representations in domains requiring factual rigor, such as physics, chemistry, or biology.

Considering these findings, we propose that text-based GenAI tools are well-suited for immediate integration into a broader range of college-level science curricula, including physics, chemistry, and biology, where conceptual clarity and verbal explanation are central to student understanding.

In contrast, graphical and video-based GenAI tools are currently best suited to creative, exploratory educational contexts, such as the speculative modelling of extraterrestrial environments in astronomy. In these settings, where imaginative engagement is prioritized over factual accuracy, visual GenAI can effectively enhance student interest, promote creative thinking, and support interdisciplinary learning without the risk of disseminating misinformation. Nevertheless, the broader application of image and video GenAI in science education, particularly where accuracy and detail are paramount, remains constrained. Progress in this area will depend on future advancements in GenAI capabilities and/or the development of targeted pedagogical frameworks that can safely integrate these tools without compromising scientific integrity.

**AUTHOR**

**Dr. Vasiliy Znamenskiy**, who holds a Ph.D. in Physics and Mathematics, teaches physics, astronomy, computer methods in science, and computer-aided analysis for engineering at CUNY (BMCC and CITY TECH). His interests include molecular dynamics simulations, quantum chemistry, hydrogen bonding, neuromorphic systems, and AI-enhanced science education. He integrates computational techniques and AI into both student learning and scientific research.